\documentclass[12pt,preprint]{aastex}
\slugcomment{Accepted by the Astrophysical Journal}
\newcommand{\water}{\hbox{H$_2$O }}

\newcommand{\arcsns}{\char'175}
\newcommand{\kms}{~km~s$^{-1}$}
\newcommand{\kmss}{~km~s$^{-1}$~}
\newcommand{\vlsr}{\hbox{V$_{lsr}$}}
\def\i16{IRAS~16342}
\begin{document}

\title{The Motion of Water Masers in the Pre-Planetary Nebula
IRAS~16342$-$3814}

\author{M. J Claussen\altaffilmark{1},\,\,R. Sahai\altaffilmark{2},\,\, 
and M. R. Morris\altaffilmark{3}}
\altaffiltext{1}{National Radio Astronomy Observatory (NRAO)
Array Operations Center, P.O. Box O, Socorro, NM 87801, USA;
mclausse@nrao.edu}
\altaffiltext{2}{Jet Propulsion Laboratory, MS183-900, California Institute
of 
Technology, Pasadena, CA 91109}
\altaffiltext{3}{Division of Astronomy and Astrophysics, University of
California
at Los Angeles, 405 Hillgard Avenue, Los Angeles, CA 90095}
\newcount\h \newcount\m \newcount\t
\h=\the\time \m=\h \divide\h by 60
\t=\h \multiply\t by 60 \advance\m by-\t
\t=\m \divide\m by 10 \multiply\m by 10 \advance\t by -\m \divide\m by 10


\begin{abstract}

We present high angular resolution observations, using the 
Very Long Baseline Array (VLBA) of the NRAO, of the high-velocity
water masers toward the ``water-fountain'' pre-planetary
nebula, IRAS~16342-3814.  The detailed structure of the water masers
appears to be that of bow shocks on either side of a highly collimated
jet.  The proper motions of the water masers are approximately equal
to the radial velocities; the three-dimensional velocities
are approximately $\pm$180\kms, which leads to a very short dynamical
time-scale of $\sim$ 100 years.  Although we do not find direct 
evidence for precession of the fast collimated jet, there may be
indirect evidence for such precession.

\end{abstract}


\section{Introduction}
It is well accepted that asymptotic giant branch (AGB) stars evolve into
planetary nebulae (PN), but the mechanisms and details are still
uncertain. What seems to be clear is that the period of transition, namely
the pre-planetary nebula (PPN) phase, is probably rather short, 
and thus observational evidence of what transpires during this interesting
phase is somewhat lacking. The development of a bipolar reflection nebula
--- two opposing lobes, usually with an equatorial ``waist'' of high
extinction --- around a star which has probably evolved off the AGB may be
a common phase leading to the establishment of asymmetric planetary
nebulae (Sahai et al. 2007). The mechanism for how this bipolarity is 
established during this evolutionary phase is still an ongoing subject 
of intense discussion (e.g., Balick \& Frank 2002). Mechanisms which have 
been considered include: (1) shaping of the outflowing wind by the 
gravitational field of a binary companion (Morris 1981, 1987; Mastrodemos 
\& Morris 1999), (2) dynamical forcing by a strong, stellar magnetic field 
(Garcia-Segura et al. 2005), or (3) sculpting of a prior, more spherically 
symmetric outflow by wandering and/or episodic jets (Sahai \& Trauger 1998; 
Soker 2002). Some combination of these mechanisms (e.g., Soker \& Rappaport 
2000) may be necessary for a complete accounting of bipolarity in particular, 
and asphericity in general, but one of the strongest constraints on the
mechanism responsible is set by the timing of its observable onset. In a
handful of sources, bipolarity appears very early in the post-AGB stage of
evolution, and in some cases when it is thought that the central
star is still on the AGB (e.g., V~Hydrae --- Kahane et al. 1996, and OH~231.8+4.2
--- e.g. Kastner et al. 1998).

Molecular masers are ubiquitous in the circumstellar envelopes of 
oxygen-rich AGB stars; in general the distribution of masers
in the AGB stage traces more or less spherically symmetric shells
of gas, with SiO masers near the stellar photosphere, water masers
further out in the shell (a few tens to a hundred A.U.), main-line
OH masers next, and finally in the far reaches of the CSE (a few
hundred to 1000 A.U.), the 1612 MHz OH masers. The total velocity spread of
these masers is set by the expansion speed of the AGB ejecta, which
typically lies in the 5---20\kms~range. 

A particularly interesting sub-class of PPNs, is the group of so-called
``water-fountain'' nebulae,  whose distinguishing characteristic is the
presence of very high velocity red- and blue-shifted \water and OH maser
features, with velocity separations in the range 50 --- 150\kms.  When examined 
with sufficient spatial resolution (using Very Long Baseline Interferometry ---
VLBI,  or the VLA), the
red and blue-shifted water masers in these sources are typically displaced 
from each other and show large proper motions, so the three dimensional (3-D) separation
velocities of the "water fountain" sources thus exceeds 80\kms, and, as
we show here, can be as high as 370\kms.  The compact size and high
velocity imply lifetimes on the order of 50 ---- 100 years, so it is not surprising
that these objects are relatively rare.
At the moment there are five confirmed members of this
type of object: IRAS~16342-3814, IRAS~19134+2131, W~43A (Likkel,
Morris, \& Maddalena 1992, hereafter LMM92), OH12.8-0.9 (Boboltz \& Marvel 2005; 2007),
and IRAS~19190+1102 (Likkel 1989; Claussen et al. 2009); although a few 
other candidates for inclusion in this class have been identified:  
IRAS~16552-3050 (Suarez et al. 2007), IRAS~18043-2116
(Deacon et al. 2007), and IRAS~18460-0151 (Deguchi et al. 2007).

In the water-fountain nebulae, the water masers appear to have been re-born
in a fast outflow. The most extreme of the water fountain nebulae is
IRAS~16342-3814 (hereafter \i16). This interesting source has water masers
spread over a range of radial velocities encompassing 270\kms, the largest
spread of velocities known in the handful of water-fountain nebulae. In the
discovery single-dish spectra (Likkel \& Morris 1988) and subsequent single-dish observations
(LMM92), the spectra show two main groups of
masers: one at radial LSR velocities from approximately 150 to 180\kms,
and one at radial LSR velocities from $-$90 to $-60$\kms. In the LMM92 multi-epoch
study, no water maser emission was found in the
intervening range of the central velocities, i.e. approximately $-$60 to $+$150\kms.
LMM92 suggested that the water masers appeared to be reflection symmetric about a
central velocity and analyzed red/blue pairs of features (e.g. $\sim +$170, $-$85; $\sim +$
145, $-$55\kms) which were thought to be 
symmetrically situated  about a single centroid velocity.  In their analysis, they
found that the central velocity was +43.2\kmss with respect to the LSR.  This central
velocity is comparable to that estimated from OH observations by Sahai et al. (1999).
LMM92 also suggested that the members of the red/blue pairs presumably arise on opposite
sides and at similar distances from the star.

\i16 has been studied in some detail in the optical and near-infrared, and
in OH maser emission. Hubble Space Telescope (HST) and Keck Adaptive Optics
(AO) images of \i16 show a small bipolar nebula, with the lobes separated
by a dark equatorial waist (Sahai et al. 1999: STMZL99; Sahai et al. 2005).  The
image morphology implies that the lobes are bubble-like reflection nebulae
illuminated by starlight escaping through polar holes in a dense, dusty
waist obscuring the central star. The AO observations reveal a remarkable
corkscrew-shaped structure apparently etched into the lobe walls, which is
inferred to be the signature of an underlying precessing jet -- this jet
has presumably carved out the observed bipolar cavities in a surrounding
AGB mass-loss envelope (Sahai et al. 2005). VLA maps of the OH maser
emission show features with the largest red- and blue-shifted velocities 
concentrated around the bright eastern and western polar lobes, respectively, 
whereas intermediate-velocity features generally occur at low latitudes, in the
dark waist region.

In this paper, we present a high angular resolution study of the water maser 
emission in \i16 with
the VLA and VLBA. VLBA studies of the water masers in the other two water
fountain PPNs, IRAS~19134+2131 and W~43A, have been carried out in the
past, yielding detailed information about the jets which are believed to
be responsible for producing the masers (Imai et al. 2002, 2004, 2007). However,
in \i16 and IRAS~19134+2131 we have the additional advantage of knowing the morphology
and orientation of the bipolar nebula. The goal of our study is to locate
the water maser features, which represent the fastest moving material
observed in this object, relative to the optical nebula, and to measure their
proper motions. Preliminary results from this study have been
presented by Sahai, Claussen, \& Morris (2003), Morris, Sahai, \&
Claussen, (2003), and Claussen, Sahai, \& Morris (2004).

\section{Observations and Data Reduction}

The observations that we describe here were performed with the
Very Long Baseline Array (VLBA) and the Very Large Array (VLA)
of the National Radio Astronomy Observatory.
The VLBA is a radio interferometer using ten 25-meter antennas (spread
across the continental United States, with two island locations
on Hawaii and the U.S. Virgin Islands (see Napier et al.  1994 for 
a full description of the VLBA).  The VLBA makes use of VLBI techniques; 
the VLA is a phase-stable, connected element interferometer, consisting 
of 27 25-m antennas in a Y-like configuration on the Plains of San Agustin 
in west-central New Mexico (reference).

\subsection{VLBA Observations}

We observed the water masers toward IRAS~16342 at a wavelength of 1.3 cm 
(rest frequency 22235.08 MHz) using the
VLBA at six different epochs:  2002 February 03, 2002 March 03, 
2002 April 04, 2002 May 05, 2002 June 03, and 2002 July 03.  
All six observations were performed in the same manner, using
two sets of 8 MHz bandwidth (each set with two circular polarizations).  
One 8 MHz set was centered at $+$155.0\kmss with respect to the local standard
of rest (LSR), and the other set was centered at $-$71.5\kmss \vlsr.
The bandwidth of 8 MHz corresponds to a total velocity range 
of 108\kmss at the frequency of the water maser line.  Thus 
the velocity coverage was from $+$209.0\kmss to $+$101\kmss and 
from $-$17.5\kmss to $-$125.5\kms.  Each 8 MHz bandwidth was correlated
with 512 spectral points across the band.  Thus each spectral
point had a velocity width of 0.211\kms.

We observed IRAS~16342 and associated calibration sources for 
five hours for each epoch.  The VLBA data were 
correlated at the VLBA correlator in Socorro, NM.  Post-processing of 
the correlated data was performed using the NRAO software package
known as Astronomical Image Processing System (AIPS; Greisen 2003).  
The correlated data were amplitude-calibrated by using {\it a priori} 
knowledge of the gains and system temperatures of the receiving 
systems / antennas provided by NRAO.  Residual delays were measured 
using the strong continuum source J1733$-$1304, which was observed 
for this purpose for several minutes approximately once per hour 
throughout each observing run. The amplitude response across the 
bandpass was calibrated, also using the observations of J1733$-$1304.  
After delay calibration, global fitting of the fringe rates of the
strong maser channels at the peak of the two maser groups was performed.
Applying the fitted solution has the effect of placing each of the 
peaks at the phase center of the subsequent images, i.e. position
information
is lost at the imaging stage.  Additionally, we applied the fringe-rate
solution of the blue-shifted peak to the red-shifted masers before mapping,
and vice-versa.  This has the effect of finding the relative position
between
the peak emission in the blue-shifted maser and the red-shifted masers
(and vice-versa).  The {\it relative} positional accuracy obtained in
this way is conservatively estimated to be 100 $\mu$as (micro-arcseconds).
For the July 2002 VLBA observations, the masers were too weak for fringe-fit
solutions to be determined. A combination of poor weather at several stations
(causing the noise level per baseline be higher) and the decline in peak flux 
density of the strongest maser features was responsible for this deficiency.
This epoch of VLBA observations is excluded from the discussion in the rest 
of this paper. 

\subsection{VLA Observations}
At each epoch we also observed the water masers with the Very
Large Array, ostensibly for the purpose of determining an 
accurate position of the masers. The goal was to provide
a reference for the VLBA observations of the masers in order
to provide accurate absolute astrometry.  Because the VLA correlator
is more limited in its capabilities (e.g number of spectral channels
and bandwidth) as compared to the VLBA correlator, the
VLA  observations were made in a mode that allowed two velocity
settings simultaneously in different polarizations.  The bandwidth
for each of the velocity settings was 3.125 MHz, yielding a 
velocity coverage for each bandwidth of about 42\kms.
In order to cover all the expected maser emission, two sets of two pairs
of velocity settings were interleaved during each set of VLA
observations.  The radial velocities of the centers of the VLA
bands were $+$172.0\kms, $+$155.0\kms, $-$55.0\kms, and $-$88.0\kmss
\vlsr.  Notice that for the $-$55.0\kmss band, the range of velocities covered
is from $-$34.0 to $-$76.0\kms, excluding the emission seen with the VLBA
at $-$25 to $-33$\kms (Section 3.2).  For the first three epochs the VLA
was used in a ``normal'' calibrator switching mode, moving between
the target and a calibrator source; spending approximately 2 minutes
on the calibrator and 10 minutes on the target.  For the final three
epochs we employed the so-called ``fast-switching'' mode, spending
approximately 70 seconds on the calibrator for every 140 seconds on
the target.  For the May and June 2002 observations, an error in setting
the frequency for the central velocity of the $-55$\kmss band was made, which
resulted in a shift of the central velocity in this band to higher negative
velocities.
IRAS~16342 is at a very low southerly declination ($\sim-$38.3$\arcdeg$)
and thus only culminates at an elevation of 20$\arcdeg$.  Due to VLA
scheduling, we didn't always observe IRAS~16342 near transit.  We have
made no corrections to the data for atmospheric opacity.  

Images were made of each spectral channel for each epoch, in order to measure 
the positions and strengths of the water maser features.  Although the rms
noise per channel varied over the 6 months of observations due to differing
weather conditions, all the observations had at least an rms noise per channel
of 20 mJy/beam.


\section{Results}

\subsection{VLA Water Masers}
We have detected water masers with the VLA for each observing session
except for the July 2002 observation, when the weather and the phase
stability of the VLA was quite poor (and the masers were likely quite
weak).  Figures 1 and 2 show spectra, made from the VLA data,  of the red-shifted
and blue-shifted water masers toward IRAS~16342 for five of the six months.  The
systemic (stellar) velocity of IRAS~16342 is +42$\pm$2\kms, based on LMM92's analysis and
observations of the OH masers (Sahai et al. 1999).  Note that the 
peak maser flux densities vary over the period of our observations, increasing until May 
2002, and then decreasing afterward.  For each epoch, the positions of the
red-shifted and the blue-shifted masers were measured.  
Three velocity complexes are apparent from
the VLA observations:  1) $+$150$\rightarrow$ $+$160\kms; 2) $+$166$\rightarrow$ 
$+$186\kms; and 3) $-$70 $\rightarrow$ $-$60\kms.  We did not detect any
masers in the velocity region near $+$145\kms, nor in the region near $-$55\kms.
These velocity ranges {\it did} have detectable masers in the study of LMM92; thus
significant changes have taken place in the maser spectra in the intervening 13 years.

\subsection{VLBA Water Maser Images}
We detected water masers using the VLBA for each observing session except for the
July 2002 observation.  Ranges of LSR velocity in which masers were detected are:
1) Figure 3 shows the general distribution of the water masers  for the epoch
of February 03, 2002 relative to the optical bipolar nebula and the OH maser emission
mapped with the VLA (from STMZL99). Since the absolute astrometry of the optical
nebula is only accurate to about 0\farcs5 or so, accurate registration of the VLBA
\water maser features to the optical image cannot be done in an absolute sense. We
have therefore assumed that the kinematic center of symmetry of the \water maser
features, C(\water), coincides with the center of symmetry of the optical nebula,
C(opt). C(\water) is determined by taking the average of the positions of the
symmetric velocity-pairs\footnote{as defined by LMM92, i.e. symmetric about the systemic
velocity} at -65\kms and 150\kmss \vlsr. 
There are four velocity groups of masers found in the February 2002 data as shown in 
Table 2.  The flux-weighted LSR velocities of these four groups are $-$66.0, $-$29.5, 
$+$178.7, and $+$153.0\kms; we 
will refer to these groups by their flux-weighted velocities in what follows.   
The maps of the masers are similar for all 5 ``good" epochs (i.e.,
excluding the July 2002 epoch).  In only one epoch, that of May 2002, is one group
(the $-$29.5\kmss group) missing.  It is clear that the groups 
of extreme velocity masers (both red and blue-shifted) are situated  
just outside the optical lobes. The $-$29.5\kmss group 
is located in the optical lobe region, near the most
blue-shifted OH masers. The extreme blue-shifted water maser regions are offset slightly
south of the lobe tip, that is, in a clockwise direction from the axis joining the central axes
of the two lobes, whereas the two groups of red-shifted water maser regions are 
offset slightly north of the lobe tip, again, clockwise from the same
axis.  Thus, the geometrical location of these regions fits in well
with the overall point-symmetric geometry of the optical lobes (classified as Bcw,ps(s)
in the morphological scheme of Sahai et al. 2007), and is consistent
with the idea that a precessing jet is responsible for producing the morphology of
this object as well as the \water masers. 

While Figure 3 shows the large-scale distribution of masers relative to the 
optical nebula, the angular resolution of the VLBA allows us to study the 
fine structure of the masers in several groups or clusters of maser sources, 
corresponding to the velocity features in the single-dish or VLA spectra.
Figures 4 - 7 show the distribution of masers on scales of tens of milliarcseconds.
In constructing these figures, we have plotted the positions of maser spots in
every frequency channel, thus, in effect, ignoring the velocity width of the
maser lines; i.e., a given maser has some intrinsic width in velocity --- since
the lines are well-resolved in velocity, plotting the position from every channel
is somewhat redundant.  The origin of these figures is the center of the line joining
the extreme velocity ``pair'': the $-$65.2\kmss and $+$154.2\kmss (with respect to the
LSR) water maser features (see below).
 
The $-$66 and $+$178.7\kms~features are spread along arc-like structures that are almost perpendicular
to the vector joining the red- and blue-shifted features.  Their 15 or 20 degree tilt with
respect to
the perpendicular is in the same sense on both sides, and is consistent with the overall point 
reflection symmetry of the nebula. The chords joining the end-points of the
largest such arcs are about 19 mas in extent. The arcs are slightly curved such that their center
of curvature lies towards the center of the nebula.  In contrast to these arc structures, the
153\kms~features for each epoch are clumped closely together (within a $\sim$4 mas region). The
$-$29.5\kms~features do not show any distinct pattern.  There is no obvious systematic velocity
change along the arcs.

Although we are unable to measure absolute positions, we can 
estimate relative proper motions of the masers with respect to one maser
feature. While not as satisfying as measuring absolute proper motions, an
advantage to measuring relative proper motions is that motions due to
parallax and motions that are in common with all the masers (secular
motion of the system, for example) are removed. Motions of water masers
toward young stellar objects have been tracked by this method (Claussen et al. 1998).
In this paper we take a slightly different tack. We  measure the
length and the position angle of a line connecting two maser spots for all five of the epochs.
For this measurement, we have selected the maser spots at the extreme ends of 
the maser distributions; viz. the maser spot with LSR velocity of $+$154.2\kmss at
the north-east extreme, and the maser spot with LSR velocity of $-$65.2\kmss at
the south-west extreme.  The length of this line changes monotonically with time
over the course of these observations, while the position angle remains constant.
The accuracy to which we measure the line length is about 200 $\mu$as, depending
slightly on epoch.  The position angle of this line is 66.1$\arcdeg$ (measured east from 
north) and is measured to an accuracy of better than 0.1$\arcdeg$.  Table 1 summarizes
this measurement for the five epochs.  The change in the length of this line, of course,
is due to maser proper motions at both ends of the line, but we cannot tell with the current
data how much proper motion to assign to either end of the line.  Thus we simply divide
the proper motion into two halves and assign one-half to each group of masers, following
the kinematical symmetry implied by the radial velocities. Since we
accurately know the relative positions of all the rest of the masers with respect to
either of the extreme ends of the distribution, assigning the proper motions this way
also fixes the motions of all the rest of the masers. 




Our analysis here and below is based on the \water maser features that form a symmetric
velocity pair about the $+$42\kmss centroid velocity: the $-$66.0 and $+$153.0\kmss groups.  
The four ``separation" proper-motion vectors (i.e., the total change in the length of the 
line described above) which we derive from our five consecutive epochs of good data
for these features, with amplitudes 1.7, 1.8, 2.4, and 1.5 mas, are consistent with
the above interpretation. In fact, it is interesting that we do not see larger
fluctuations in the proper motion over time, suggesting that the distribution of
density and temperature in the jet head are maintained over at least several month
time-scales.  A linear least-squares fit to these amplitudes with time gives a formal
estimate of 63$\pm$2 $\mu$as day$^{-1}$ (see Figure 8).

We henceforth assume that $\mu=63\,\, \mu$as day$^{-1}$ (or 23.0 mas yr$^{-1}$ for the proper motion
of the ``separation vector'') represents the proper motion of these two groups of \water masers 
due to their physical 3-D motion.  Apportioning one-half of the proper 
motion to each group of masers at the ends of the line results in a tangential velocity for each group 
(i.e., the component of 3-D velocity in the sky-plane), $V_t=\mu d=109.8$\kms, where 
$d$ is the distance (assumed to be 2\,kpc from STMZL99).  Since the line-of-sight velocity is
$V_r=109.8$\,\kms, we obtain an inclination angle, $i=tan^{-1}(V_t/V_r)=45.0\arcdeg$
(where $i$ is the inclination of the nebular axis to the line-of-sight), which is in
excellent agreement with the independent estimate by STMZL99, who derived
$i\sim40\arcdeg$ (by assuming that the observed velocity spread in the 1612 MHz
emission features clustered around the base of each lobe is comparable to the
difference in projected radial velocities of the front and back sides of the lobe).
The total 3-D speed (one-sided) of material in the jet head, based on the $-$66.0\kmss and $+$153.0\kmss
\vlsr velocity pair, is thus 155.3\,\kms. This is very similar to the 3-D jet speeds
inferred for the other water fountain PPNs, W43A (150\kms: Imai et al. 2002) and
IRAS19134 (130\kms: Imai et al. 2004).

We can also perform a similar analysis relating the $+$178.7 group of masers relative
to the $-$66.0 group (though these two groups do {\it not} form a symmetric pair). 
The four ``separation'' proper-motion vectors between the $+$178.7 and
the $-$66.0\kmss groups have amplitudes 2.1, 1.7, 1.7, and 2.9 mas.  A linear least-square
fit to these amplitudes over time gives 67$\pm$5 $\mu$as day$^{-1}$, slightly higher than
that derived from the $-$66.0 and $153.0$ groups.  Thus the proper motion 
of the $+$178.7 group appears to be somewhat higher than that of the $+$153.0 velocity group.  
This value gives 13.0 mas yr$^{-1}$ for the proper motion of the $+$178.7 velocity group 
(assuming that the $-$66.0 group has a proper motion of 31.5 $\mu$as day$^{-1}$).  Thus the tangential 
velocity, the inclination angle, and the total 3-D speed of material in the $+$178.7\kms 
velocity group are 134.2\kms, 42.8$\arcdeg$, and 182.8\kms, respectively.   It is interesting
to note that the difference in position between the April, May, and June observations is not
very uniform; we suggest that this is due to a clumpy medium through which the jet is traveling.

We can derive a distance-independent age for the jet in \i16 by dividing the
projected (on the sky-plane) separation of the masers by the expansion proper motion. 
For the $-$66.0 and $+$153.0\kmss pair, the age estimate is $\sim$130\,yr, assuming that the average
proper motion has remained constant.  For the $+$178.7\kmss group the age estimate is $\sim$107\,yr.
However, since the speed of the underlying jet which is pushing and compressing the \water 
emitting material must be greater than the speed of the latter, these ages are upper limits. 
Making a similar assumption of constant proper motion, the ages of the jets in W43A and
IRAS19134, 
respectively, are 35 and 50 yr (Imai et al. 2004), significantly smaller than that in \i16. This
may explain why only \i16, amongst the three water-fountain PPNs, has a well
developed optical nebula.

Comparing the velocities of the maser features from our data set with the data sets
presented by LMM92, we find a systematic increase in the velocity offset of the
blue- and red-shifted features from the systemic velocity over the time period from
1987/89 to 2002 that are in common in the two data sets. In LMM92's data, there are 
three sets of closely spaced velocity (see Table 2)
features (velocity complexes) which have counterparts in our data. In the LMM92 data,
these are the $-$60.5, the $+$147.7, and the $+$172.7\kmss groups, which are matched
in our data by the $-$66.0, $+$153.0, and $+$178.7\kmss velocity groups.
We have computed the peak-intensity-weighted average velocities of these
complexes, both from the LMM92 data and ours; in comparing these we
find that the outflow speed of
the maser features has increased roughly by about 5-6\kms~over a period of about
$\sim$13 --- 14 years (we use only the 1988 and 1989 data from LMM92). Thus, during this period, 
momentum has been added to the \water maser
emitting material by the jet, implying that the jet in \i16 has remained active. The
increase in the outflow speed may be due to an increase in the momentum of the jet,
or to a decrease in the resistive force of the ambient circumstellar material into which the jet is
presumably expanding.  An estimate of the average acceleration is $\sim$0.4\kmss yr$^{-1}$.  This
compares favorably with the three-dimensional acceleration of water masers in OH12.8$-$0.9
determined by Boboltz \& Marvel (2007).

We have looked for a similar effect in W43A and IRAS19134 by comparing the maser
velocities in LMM92 with those published by Imai et al. (2002, 2004); unlike \i16, we
find no evidence for a systematic change in the outflow
velocities for these sources.

\section{Discussion}

Proper motion of the \water masers can result from either actual physical motion of
the emitting material, or to a ``theatre marquis" phenomenon, in which dense
material at progressively larger distances ``lights up" sequentially in \water maser
emission due to, for example, the passage of a shock wave that produces an outward moving
compression front. 
Numerical simulations of fast collimated jets expanding in AGB
circumstellar envelopes (Lee \& Sahai 2003) show the presence of dense material at
the head of the jet, where the physical conditions for the excitation of \water
masers (temperature $\sim$ 400\,K, density $\sim 10^9$ cm$^{-3}$: Elitzur 1992) are
likely to be maintained. The forward (bow) shock along the jet axis is located within
this dense region at the head of the jet. 
The bulk of the gas in this region is material from the slowly-expanding 
ambient circumstellar envelope that has been swept-up, compressed, and accelerated 
by a fast, tenuous (i.e., underdense with respect to the ambient medium) jet-like 
outflow.  Hence we expect that, on the average, the proper motions of the \water 
masers reflects the physical motion of the dense gas in the jet head, and this motion 
will generally be at speeds less than the intrinsic speed of the underlying jet. 
The ratio of the speed of the dense region to that of the jet depends on the
density contrast between the jet and the ambient medium (see, e.g., Eqn. 9 of 
Lee \& Sahai 2003). 
Fluctuations from the average motion may arise as the instantaneous
masing region, whose size is small compared to the size of the dense region in the
jet head, may be located at different positions within the latter at different
times. 
If, in addition, the jet is precessing, it impacts material in progressively different
directions, and one might observe the projection of this progression (another 
potential manifestation of the theatre marquis effect), which, in general, would 
consist of both radial and tangential components.  We do not observe any deviations 
from purely radial motion (i.e., motion parallel to the jet) of the maser spots that 
we have followed, so any precession of the jet that might be occurring is apparently 
not  affecting those spots.  Rather, precession of the jet may simply produce new 
maser spots in the direction of the precession.  This might account for the disappearance 
of maser features at the south end of the $-$66\kmss arc and their appearance at the
north end of this arc (Fig. 4).

The maser spot evolution diagrams shown in Figures 4 and 6 show a 
potential additional phenomenon: a slight divergence of the spot motions,   
relative to the average proper motion vector (shown as dashed lines in each 
figure).  The southernmost extreme spot group that can be followed through 
all epochs shows a directional deviation that corresponds to such a divergence.  
The significance of this phenomenon is marginal, but if it is found to 
recur in future observations, it could provide an interesting clue to the 
curvature-related dynamics of the shock front.

The distribution of \water maser features which we have found in \i16 is
generally tangential to the inferred jet axis, in contrast to that in
W43A, where the features appear to lie along the curved trajectory of
material ejected in a precessing jet. These differences may be related to
differences in the intrinsic properties of the jets or the ambient
circumstellar environment provided by the slowly expanding mass-loss
envelopes formed during the progenitor AGB phase, with which the fast jet
must presumably interact. The IRAS far-infrared fluxes from these objects
are a probe of the dense, dusty circumstellar environment in these
objects. Both objects are believed to lie at comparable distances
(2.6\,kpc for W43A and 2\,kpc for \i16), and their 12 and 25 \micron~fluxes
are rather similar (23.7 and 103.5\,Jy in W43A; 16.2 and 199.8\,Jy in
\i16) hence there is no obvious reason to believe that their circumstellar
environments are very different.

The lobes seen in the near-infrared AO images are more extended radially than the optical
lobes, and, because of the smaller optical depths in the near-IR, more accurately
delineate the true physical extent of the bipolar cavities which which are manifested as
the lobes in the images. The extreme \water masers, separated by 2.978\arcsns, lie near 
(but not exactly) at the tips of the near-IR lobes, thus near the ends of the bipolar 
cavities. The point-symmetrically distributed offsets of these features from the tips 
of the infrared lobes is to be noted, and suggests that the current working-surfaces of 
the bipolar jet have moved away from the physical ends of these elongated cavities, 
providing further support for the idea that the cavities are being carved out
by a precessing bipolar jet.

The $-$29.5\kms~masers do not show obvious evidence for systematic proper motion. 
In contrast to the extreme velocity masers, these are located near the most 
blue-shifted OH masers in the western optical lobe. A comparison with the 
$L_p$ (3.8 \micron) image in Sahai et al. (2005), shows that this
maser complex coincides with the bright region labeled knot W2,
apparently part of the corkscrew structure. The associated \water emitting
material clearly does not show any systematic motion like the extreme
velocity masers (Fig. 7). An inspection of the jet-envelope interaction
models of Lee \& Sahai (2003) shows that lower outflow velocities are
generally expected for the dense swept-up material at locations on the
walls of the cavities which are at lower latitudes, compared to the
material at the tips. Thus both the proper motion and the radial
velocity of this material is expected to be less than that of the
material at the tips of the lobes because its intrinsic, 3-dimensional
outflow speed is less than that of the latter. However, the expected
proper motion of the $-$29.5\kms, assuming the same
inclination angle as for the extreme velocity masers, is about 7 mas yr~$^{-1}$, 
so it is somewhat surprising that we do not detect this in our data. We
conclude that, for these masers (unlike the extreme velocity ones), the
location of the individual masing regions changes by distances
comparable to the proper motion ($\sim$1 AU) on time-scales of $\sim$1
month. Since maser emission requires velocity coherence, and \water maser
excitation is believed to be collisionally-driven, such changes imply
that there are much larger changes in the velocity, density,
temperature and structure in region W2 than at the tips of the lobes,
where the extreme velocity masers are produced over month-long
time-scales.


The \water masers associated with the blueshifted, near lobe of the nebula, are
arranged spatially so that the highest velocity component is farthest from the
center of the system.  This could be understood simply in terms of the displacement 
being proportional to velocity.  However, this is not the case for the masers 
associated with the redshifted lobe, where the $+$153\kmss feature is located 
farther out than the $+$179\kmss feature.  Because all of the masers lie close 
to a single line joining the most distant spot groups, this is unlikely to be 
ascribable to purely geometrical effects in which the different velocities in 
either lobe result from different line-of-sight projections of jet bursts coming
out with a constant velocity in different directions at different times.  If the 
successive bidirectional jet bursts have the same initial velocities, then our 
data suggest that they have variable histories, with some being decelerated by 
direct interaction with the slow AGB wind, while others may follow the trajectories 
of previous spurts which may have cleared out some of the slow wind material. 
Of course, it is also possible that the jet bursts have variable initial velocities. 


\section{Conclusions}

We have observed the ``water fountain'' masers from IRAS~16342$-$3814
with the VLBA at very high angular resolution over a total time span of 
four months at approximately 1 month intervals. These observations show 
that the detailed structure of the high radial velocity water masers
(making up the ``fountain'') lie on opposite sides of the optical nebula.
These structures appear to be bow shocks which are likely formed at the
impact point of a highly collimated jet with dense molecular gas found
along the jet axis.  Based on our VLBA and VLA observations, compared
with earlier single-dish observations of the water masers from \i16,
we find that the ``paired'' high-velocity features are not always apparent.

The proper motion measurements, when combined with the radial velocity
measurements of the water maser spots, give a three-dimensional picture
of the kinematics of the gas.  We calculate that the 3-D velocities of 
the jet at the position of the extreme velocity maser components are at
least 155 to 180\kms.  We also calculate that the inclination of the jet 
axis to the line of sight is about 44\arcdeg, in excellent agreement with
other estimates.  Upper limits on the dynamical age of the jet (which is
distance-independent) are 110 - 130 years.

Direct evidence for precession of the fast collimated jet is lacking in
our data;  the maser features do not trace a curved trajectory as they
do in the case of W43A.  However, there may be some indirect indication 
of precession from these observations, e.g. the appearance and disappearance
of features at the ends of the arcs, as well as the difference in position
between the extreme \water masers and the tips of the infrared lobes.

Finally, there are maser features near (but not {\it} at) the systemic velocity 
which show no systematic proper motion over the four months. We 
conclude that, for these water masers, the location of such  masing regions
changes in a non-systematic manner by distances similar to the proper motion 
(~1 AU) on one-month time scales.

\acknowledgments

RS and MM thank NASA for partially funding this work by NASA LTSA award (nos.
NMO710651/ 399-20-40-06); RS also received partial support for this work
from HST/GO awards (nos. GO-09463.01-A and GO-09801.01-A) from the Space
Telescope Science Institute (operated by the Association of Universities
for Research in Astronomy, under NASA contract NAS5-26555). RS's
contribution to the research described in this publication was carried
out at the Jet Propulsion Laboratory, California Institute of
Technology, under a contract with NASA.

\clearpage
\begin{figure}
\plotone{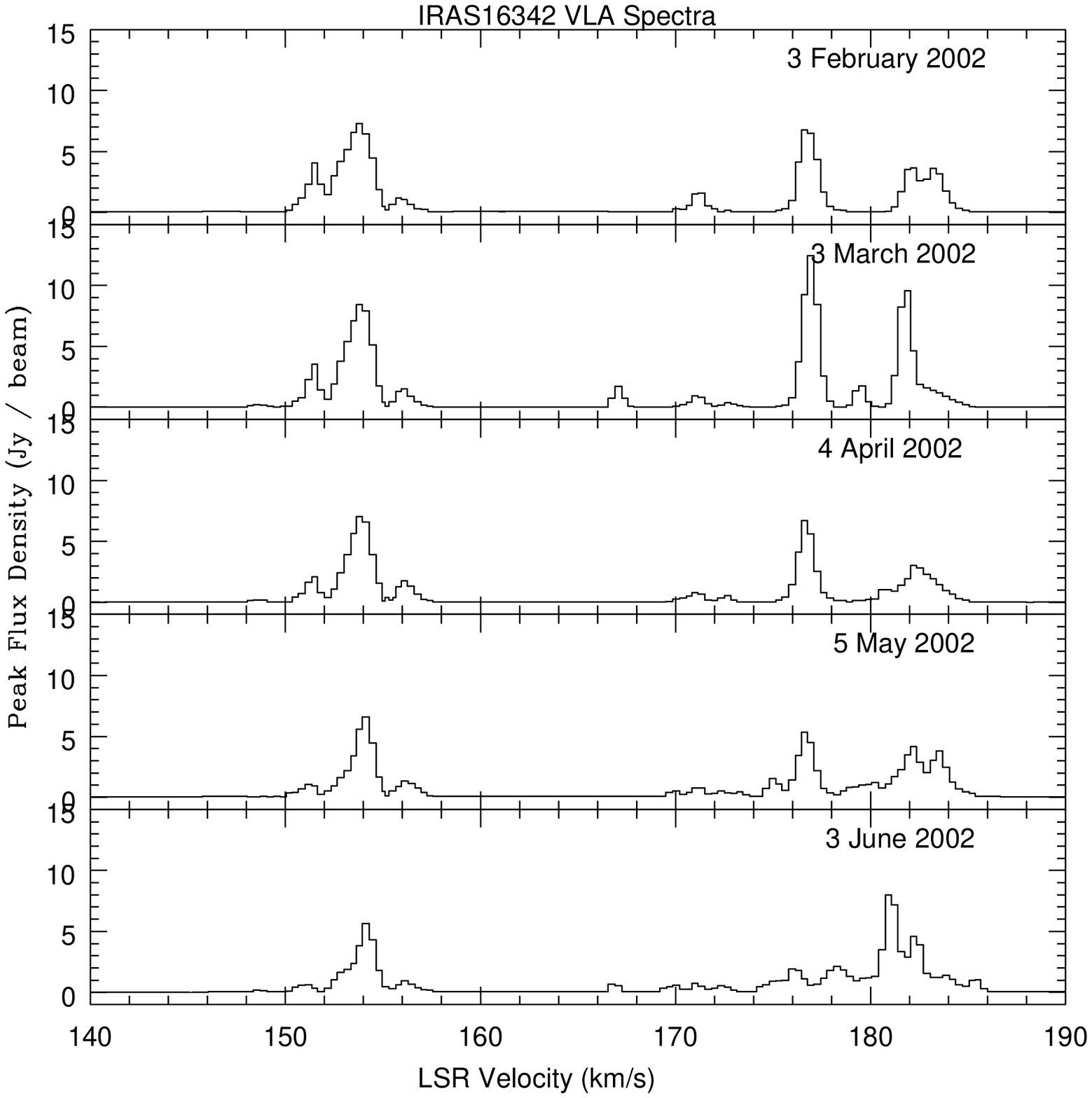}
\caption{Spectra of the red-shifted water masers observed with the
VLA for the five monthly epochs described in the text.  Each of these
spectra is comprised of two frequency settings at the VLA; the velocity
range of the plot is the entire coverage of the red-shifted masers observed
with the VLA.}
\label{red-spectra}
\end{figure}
\clearpage

\begin{figure}
\plotone{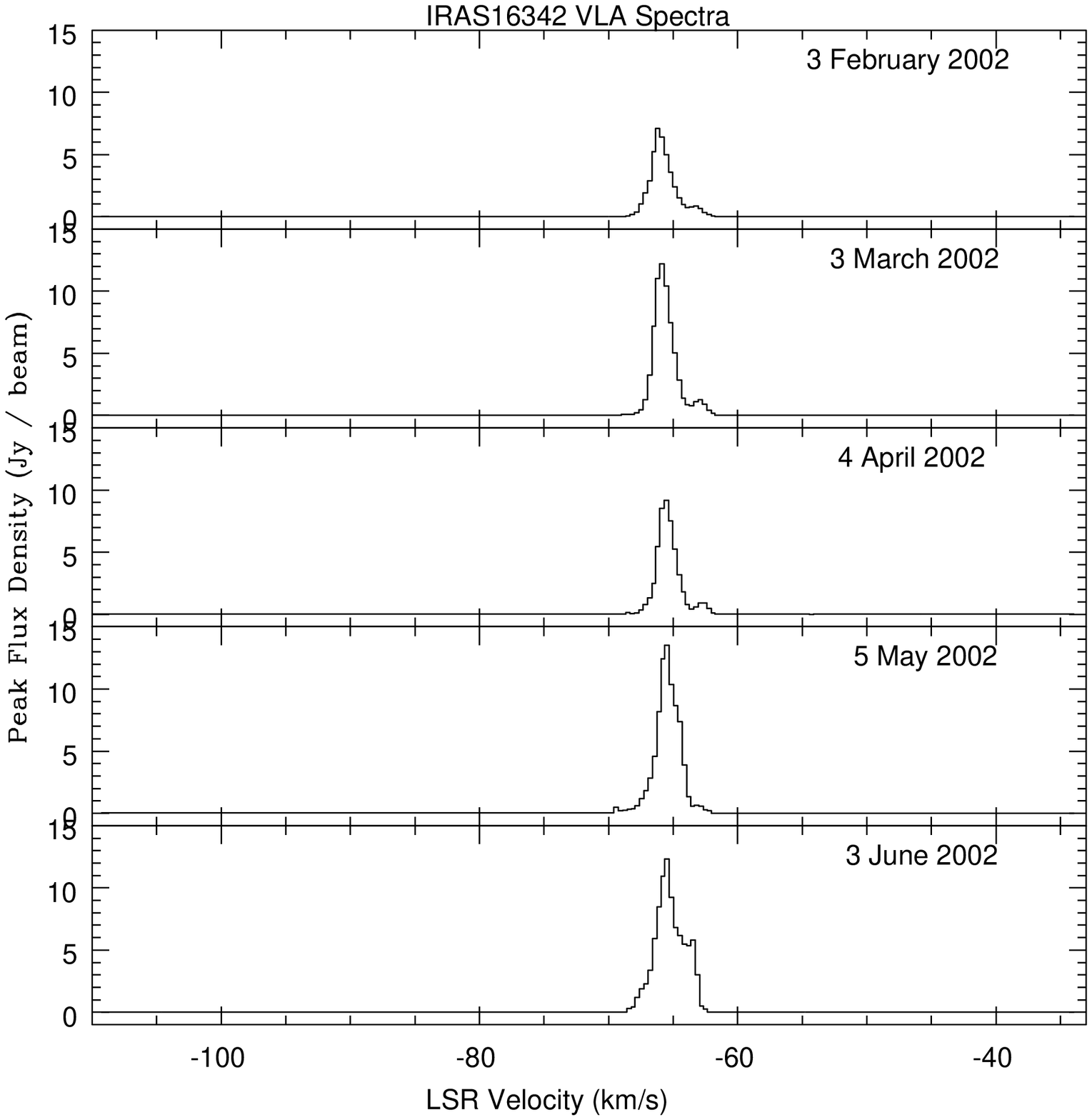}
\caption{Spectra of the blue-shifted water masers observed with the
VLA for the five monthly epochs described in the text.  Each of these
spectra is comprised of two frequency settings at the VLA; the velocity
range of the plot is the entire coverage of the blue-shifted masers
observed with the VLA.}
\label{blue-spectra}
\end{figure}

\begin{figure}[htb]
\vskip -0.9 truein
\plotone{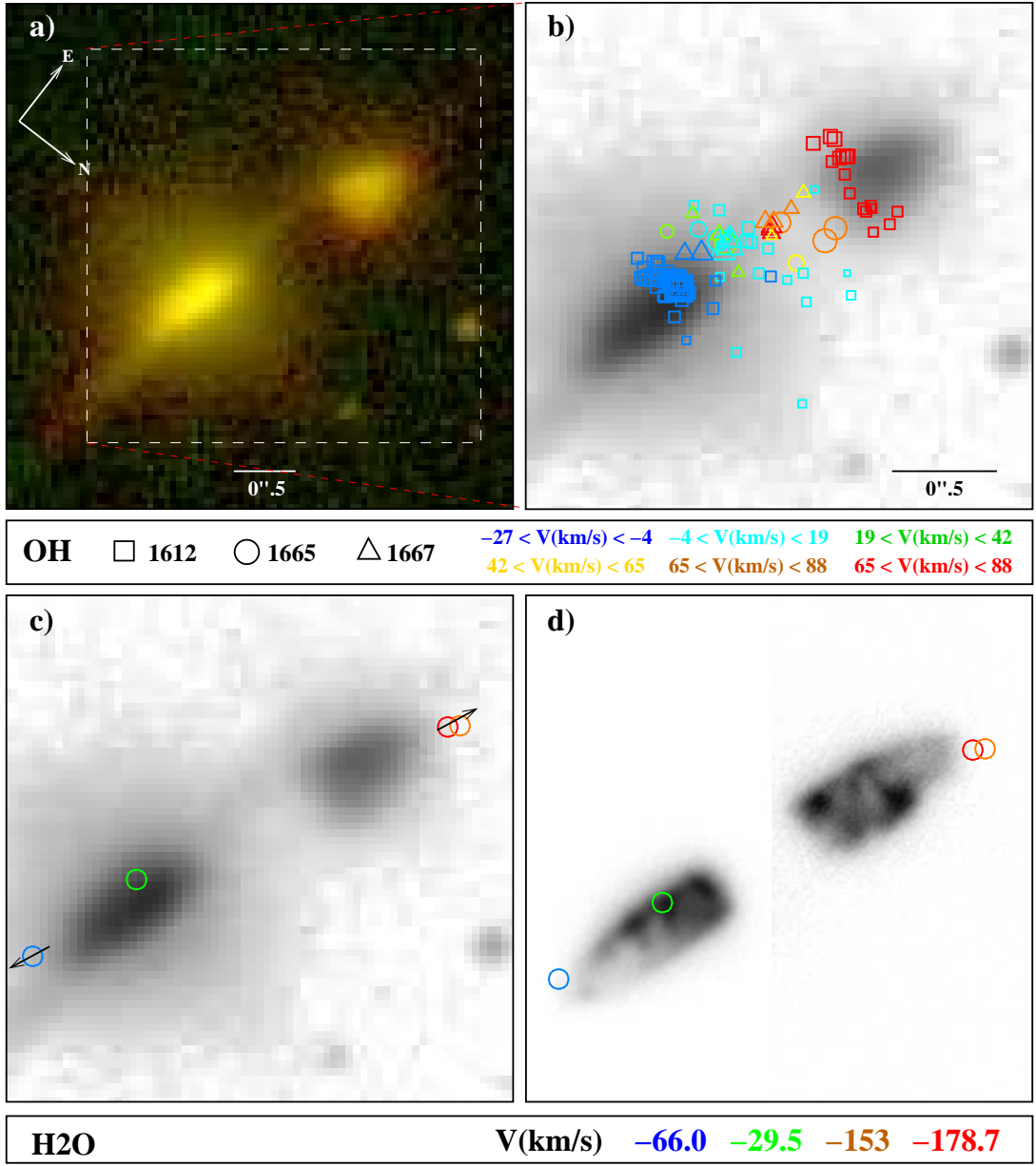}
\vskip -1.0 truein
\caption{
(a) A color image of the protoplanetary nebula IRAS16342-3814, made of exposures taken with
WFPC2/HST through the F814W (in red) and F555W (in green) filters, and displayed on a
logarithmic stretch (b) OH 1612, 1665, and 1667 MHz maser line features observed with the VLA ,
overlaid on the F814W image (in grey-scale). The maser features have been grouped into 6
color-coded velocity bins. The symbol sizes are proportional to the square-root of the
integrated flux, and the highest flux features have strengths of 8.83, 0.53, and 1.17 Jy
for the 1.612, 1.665, and 1.667 features, respectively. (c) VLBA \water 22.235 GHz features
observed with the VLBA, overlaid on the F814W image. Arrows representing the proper motion
vectors of the \water maser features are also shown; the orientation of this vector remains
fixed at $66\arcdeg.1$ over our monitoring period (Jan--May 2002). (d) \water maser features
overlaid on the $L_p$-band (3.8\micron) image, taken using Adaptive Optics on the Keck II
telescope (the $L_p$ image is adapted from Sahai et al. 2005).
}
\label{hst-oh-h2o}
\end{figure}

\clearpage
\begin{figure}
\plotone{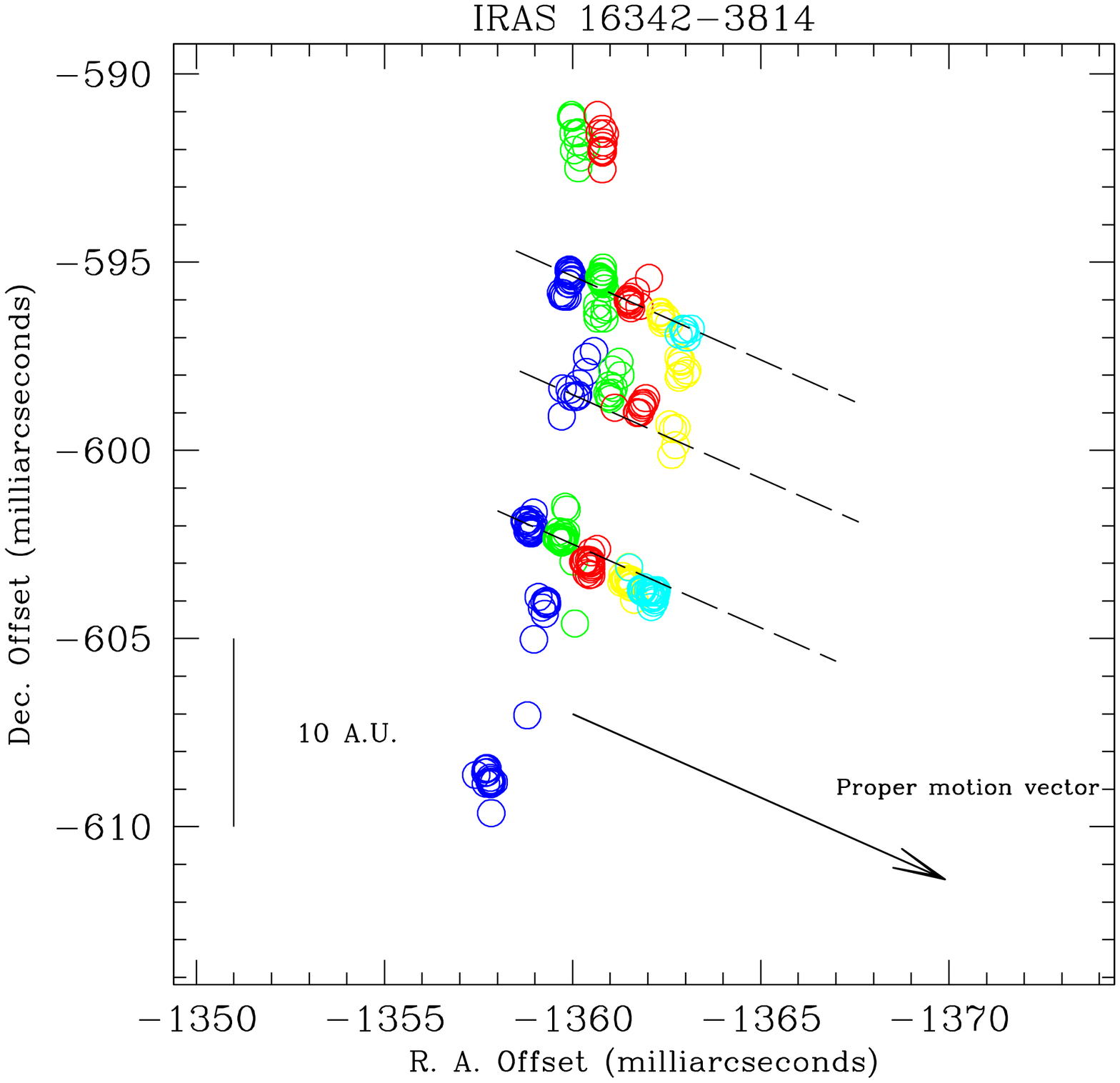}
\caption{Zoom-in of the VLBA distribution of masers for the blue-shifted ($-$66\kmss 
radial velocity features) for all five epochs. This figure zooms in on
the masers in the extreme lower left of Figure \ref{hst-oh-h2o} (i.e. outside the
southwest optical lobe). The color code represents different epochs of observation:
blue --- 3 Feb 2002; green --- 3 Mar 2002; red --- 4 April 2002; yellow --- 5 May
2002; and cyan --- 3 June 2002. All features in all spectral channels comprising
this range are plotted as open circles centered on their position. In the lower-left
hand corner is a bar showing a 10 AU extent, assuming a distance to the source of 2
kpc. Each of the dashed lines connects maser features at different epochs which we
believe arise in the same physical clump. The arrow shows a 10.8 mas yr$^{-1}$ proper
motion vector. The orientation of the axes here is different than that in Fig. 2.
}
\label{blue-distr}
\end{figure}

\clearpage
\begin{figure}
\plotone{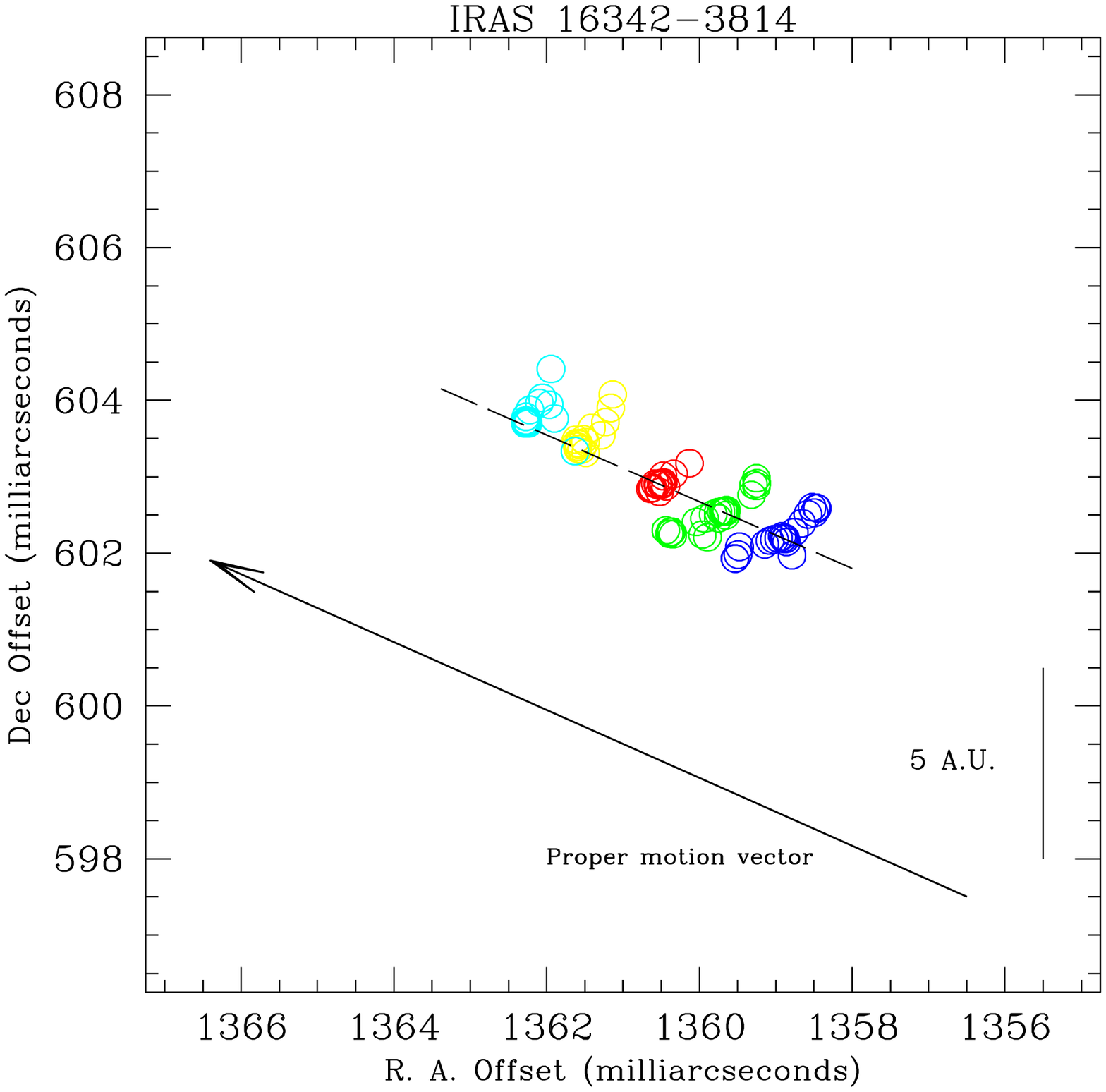}
\caption{Zoom-in of the VLBA distribution of masers for the red-shifted
($+$153\kmss radial velocity features) for all five epochs. This figure zooms in on the
masers in the extreme upper right of Figure \ref{hst-oh-h2o} (i.e. outside the
northeast optical lobe). The color code represents different epochs of observation:
blue --- 3 Feb 2002; green --- 3 Mar 2002; red --- 4 April 2002; yellow --- 5 May
2002; and cyan --- 3 June 2002. All features in all spectral channels comprising
this range are plotted as open circles centered on their position. In the lower-left
hand corner is a bar showing a 10 AU extent, assuming a distance to the source of 2
kpc.
}
\label{red155-distr}
\end{figure}

\clearpage
\begin{figure}
\plotone{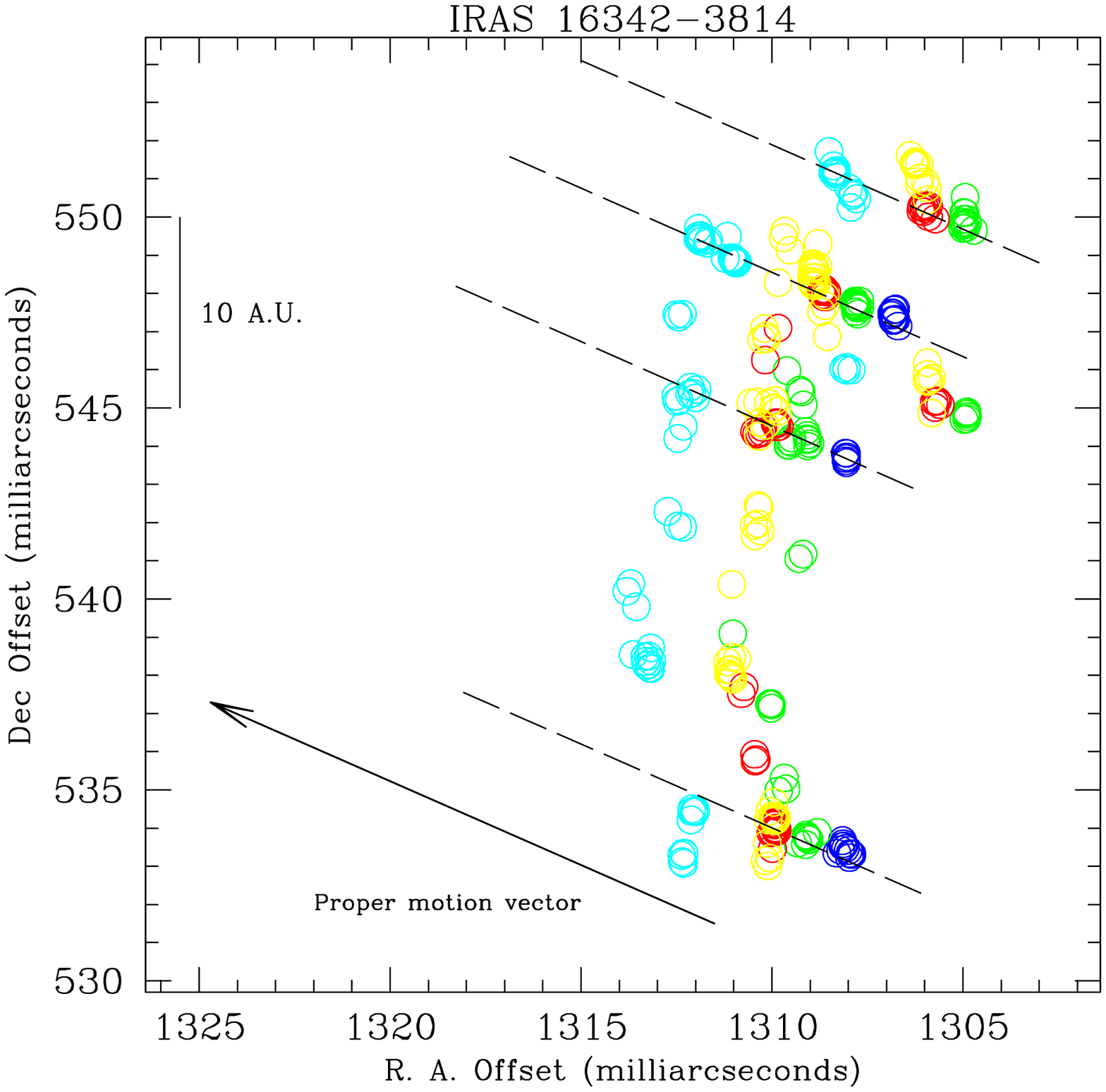}
\caption{Zoom-in of the VLBA distribution of masers for the red-shifted
($+$178\kmss radial velocity features) for all five epochs. This figure zooms in on the
masers in the upper right of Figure \ref{hst-oh-h2o} (i.e. outside the
northeast optical lobe). The color code represents different epochs of observation:
blue --- 3 Feb 2002; green --- 3 Mar 2002; red --- 4 April 2002; yellow --- 5 May
2002; and cyan --- 3 June 2002. All features in all spectral channels comprising
this range are plotted as open circles centered on their position. In the lower-left
hand corner is a bar showing a 10 AU extent, assuming a distance to the source of 2
kpc.  The proper motion vector here represents 14.4 mas yr${-1}$.
}
\label{red180-distr}
\end{figure}

\clearpage
\begin{figure}
\plotone{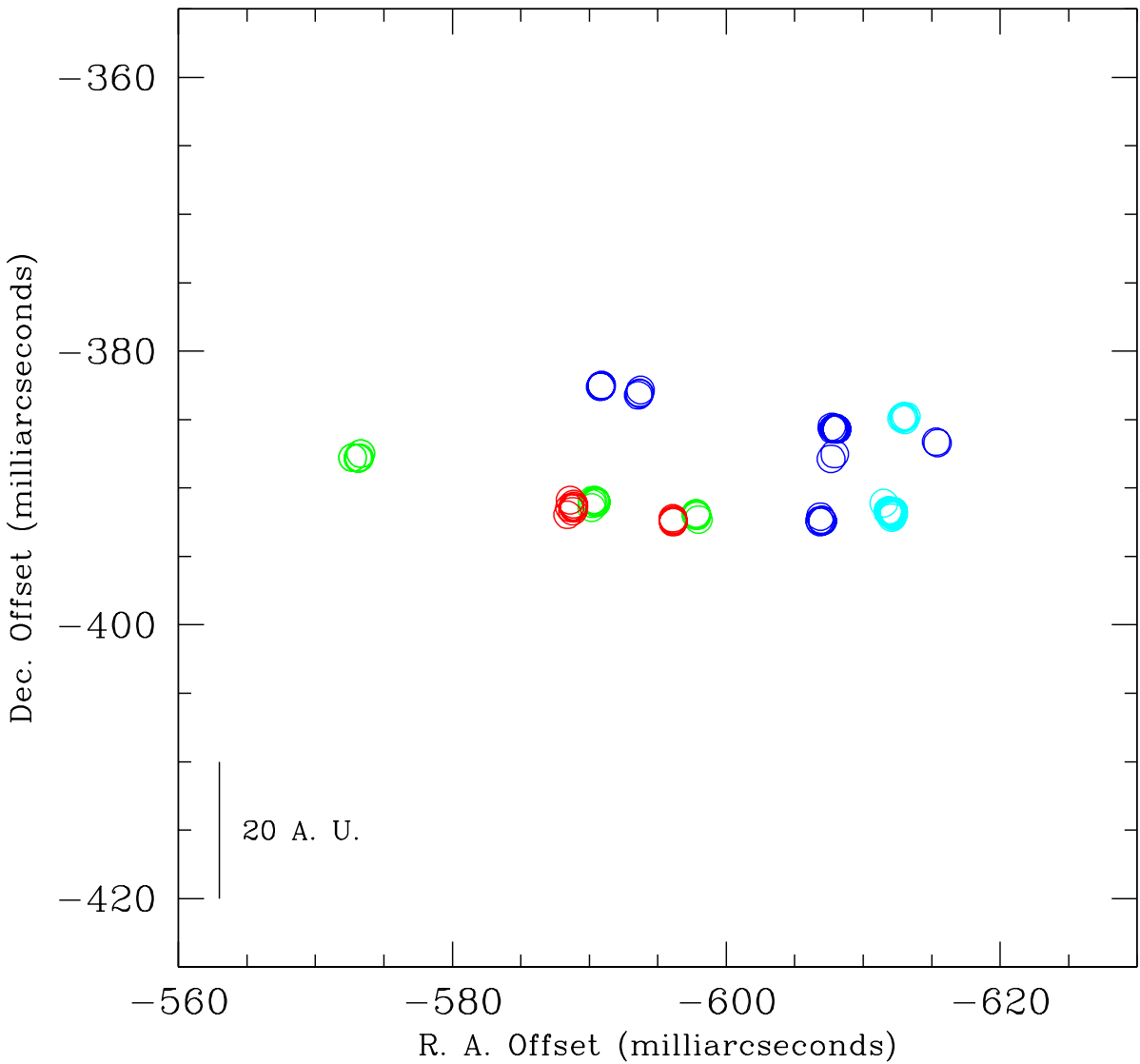}
\caption{Zoom-in of the VLBA distribution of masers for the velocity range $-$21 $\rightarrow$
$-33$ \kmss for four epochs.  The color code represents the different epochs of observation:
blue --- 3 Feb 2002; green --- 3 Mar 2002; red --- 4 April 2002; 
and cyan --- 3 June 2002. We did not detect emission from this velocity group in May 2002.
All features in all spectral channels comprising this range are plotted as open circles centered 
on their position.   Note that there is no apparent systematic motions of these masers as
compared with the those at other radial velocities in Figures 4, 5, \& 6.
In the lower-left
hand corner is a bar showing a 20 AU extent, assuming a distance to the source of 2
 kpc.
}
\label{green-33-distr}
\end{figure}

\clearpage
\begin{figure}
\plotone{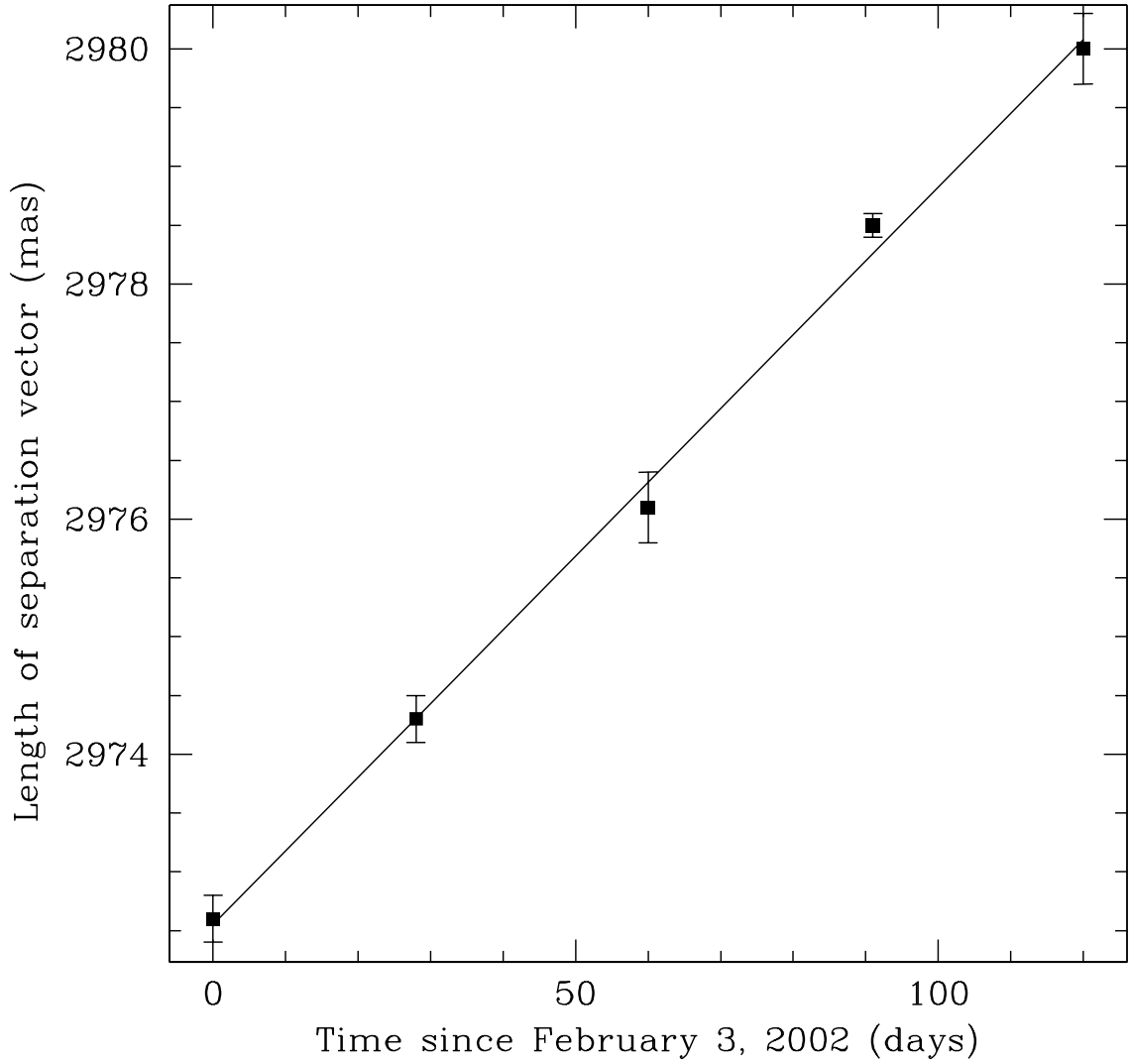}
\caption { A plot of the length of the line joining the $+$154.2\kmss feature
and the $-$65.2\kmss feature as a function of time for 5 VLBA epochs.  The line
is a linear least-squares fit to the data, giving a slope of 63$\pm$2 $\mu$as
day$^{-1}$. }
\label{pm-linelength}
\end{figure}

\begin{deluxetable}{ccc}
\tablecolumns{3}
\tablewidth{0pc}
\tablecaption{Summary of VLBA Observations}
\tablehead{
\colhead{Date} & \colhead{Length of Line from} & \colhead{Position Angle (N$\rightarrow$E)} \\
 & \colhead{$-$65.2 and $154.2$\kmss Features (mas)} & \colhead{of this Line (degrees)} \\
}
\startdata
2002 February 03 & 2972.6$\pm$0.2 & 66.1$\pm$0.04  \\
2002 March 03    & 2974.3$\pm$0.2 & 66.1$\pm$0.04  \\
2002 April 04    & 2976.1$\pm$0.3 & 66.1$\pm$0.05  \\
2002 May 05      & 2978.5$\pm$0.1 & 66.1$\pm$0.04  \\
2002 June 03     & 2980.0$\pm$0.3 & 66.1$\pm$0.05  \\
\enddata
\end{deluxetable}

\clearpage
\begin{deluxetable}{cccc}
\tablecolumns{4}
\tablewidth{0pc}
\tablecaption{Nomenclature for Flux-Density Weighted Velocity Groups }
\tablehead{
\multicolumn{2}{c}{This Paper} & \multicolumn{2}{c}{Likkel et al. (1992)} \\
\colhead {Name of Velocity Group} & \colhead{LSR Velocity Range} 
 & \colhead {Name of Velocity Group} & \colhead{LSR Velocity Range}
}
\startdata
$-$66.0 & $-$62.2$\,\rightarrow$$\,-$68.3 & $-$60.5 & ---  \\
$+$155.3 & $+$150.2$\,\rightarrow$$+\,$155.2 & $+$147.7 & $+$147.6$\,\rightarrow$$\,+$147.8 \\
---     & ---                          & $-$56.5  & $-$56.2$\,\rightarrow$$\,-$55.8 \\
$+178.7$ & $+$170.6$\,\rightarrow$$\,+$184.3 & $+$172.7 & $+$170.2$\,\rightarrow$$\,+$175.1 \\
$-$29.5 & $-$20.5$\,\rightarrow$$\,-$33.4 & --- & --- \\
---     & ---                          & $-$87.7  & $-$86.4$\,\rightarrow$$\,-$89.0 \\
\enddata
\end{deluxetable}


\end{document}